\documentclass{elsart}
\usepackage{epsfig}
\usepackage{rotating}
\begin{document}
\runauthor{Marco Battaglia}
\begin{frontmatter}
\title{A Rad-hard CMOS Active Pixel Sensor\\ for Electron Microscopy}
\author[UCB,LBNL]{Marco Battaglia\corauthref{cor}},
\corauth[cor]{Corresponding author, Address: Lawrence Berkeley National Laboratory, 
Berkeley, CA 94720, USA, Telephone: +1 510 486 7029.} 
\ead{MBattaglia@lbl.gov}
\author[LBNL]{Devis Contarato},
\author[LBNL]{Peter Denes},
\author[LBNL]{Dionisio Doering},
\author[LBNL,INFN,Padova]{Piero Giubilato},
\author[LBNL]{Tae Sung Kim},
\author[INFN,Padova]{Serena Mattiazzo},
\author[LBNL]{Velimir Radmilovic},
\author[UCB,LBNL]{Sarah Zalusky}
\address[UCB]{Department of Physics, University of California at 
Berkeley, CA 94720, USA}
\address[LBNL]{Lawrence Berkeley National Laboratory, 
Berkeley, CA 94720, USA}
\address[INFN]{INFN, Sezione di Padova, I-35131, Italy}
\address[Padova]{Dipartimento di Fisica, Universit\`a degli Studi, 
Padova, I-35131, Italy}
\begin{abstract}
Monolithic CMOS pixel sensors offer unprecedented opportunities for 
fast nano-imaging through direct electron detection in transmission 
electron microscopy. We present the design and a full characterisation 
of a CMOS pixel test structure able to withstand doses in excess of
1~MRad. Data collected with electron beams at various energies of interest
in electron microscopy are compared to predictions of simulation and 
to 1.5~GeV electron data to disentagle the effect of multiple scattering.
The point spread function measured with 300~keV electrons is 
(8.1 $\pm$ 1.6)~$\mu$m for 10~$\mu$m pixel and (10.9 $\pm$ 2.3)~$\mu$m 
for 20~$\mu$m pixels, respectively, which agrees well with the values 
of 8.4~$\mu$m and 10.5~$\mu$m predicted by our simulation.
 
\end{abstract}
\begin{keyword}
Monolithic active pixel sensor, Transmission Electron Microscopy;
\end{keyword}
\end{frontmatter}

\typeout{SET RUN AUTHOR to \@runauthor}


\section{Introduction}

\vspace{0.5cm}

CMOS monolithic active pixel sensors were first proposed as photo-detectors 
forty years ago. About a decade ago their use as charged particle detectors 
started to be explored, recognising that they represent an appealing technology 
offering small pixel size and the possibility of high speed read-out~\cite{fossum,cmos}. 
In addition to their use in charged particle tracking, they have been proposed 
and investigated for direct electron detection in electron microscopy,
in particular in Transmission Electron Microscopy (TEM) as electron 
imagers~\cite{emicro,deptuch,Denes:2007} together with hybrid 
pixels~\cite{fan,faruqi,faruqi2}, as an advantageous alternative to CCDs 
optically coupled to phosphor plates (for a review of different detection 
technologies, see~\cite{review}).

For single particle tracking, it has been demonstrated that CMOS pixels of order 
10$\times$10~$\mu$m$^2$ can reconstruct the position of impact of the particle 
with an accuracy of ${\cal{O}}\mathrm{(1~\mu m)}$, by using the centre of gravity 
of the charge distribution~\cite{mimosa}. Owing to their thin sensitive region, 
confined to the $\simeq$10~$\mu$m lightly-doped epitaxial layer, these sensors can be 
back-thinned down to $\le$~50~$\mu$m, without loss in performance~\cite{Battaglia:2006tf}.
The thin charge-collection region and total thickness of CMOS pixel sensors is ideal
in TEM applications, where multiple scattering effects are significant, for obtaining 
a good Point Spread Function (PSF). In the energy region of interest in TEM, which is 
$\simeq$ 60 - 400~keV, the specific energy loss, ${\mathrm{dE/dx}}$, varies by an order 
of magnitude but remains large enough to ensure single electron detection with large 
signal-to-noise ratio (S/N) values. For TEM, the detector should record the position 
of a collection of electrons that represents a magnified image. Whereas diffusion is 
beneficial for tracking applications, as it improves the determination of the 
charge centre of gravity,  and thus of the particle impact point, to an accuracy 
significantly better than the pixel pitch, for TEM diffusion is harmful, as it 
increases the PSF and needs to be carefully evaluated.

While significant progress has been made in the last few years to develop CMOS
pixel sensors for tracking and vertexing applications in particle collider 
experiments, several issues, relevant to their application for fast nano-imaging at 
TEM, remain to be addressed. In particular, their radiation hardness is a key 
challenge in the design of a detector for TEM. Unlike HEP, 
where the total dose requirement can be estimated, based on integrated luminosity, 
TEM requirements depend on the mode of operation. Whereas the radiation hardness
for the specimen can be easily specified (in electrons per square angstrom), as 
the magnification is variable, this does not translate into a requirement on the 
detector. In addition, large numbers of electrons may be contained in a limited number 
of Bragg spots, when operating in diffraction mode, causing large doses to be received 
by a small number of pixels. Given that a very intense bright field image could deposit 
order of 10~rad~s$^{-1}$ pixel$^{-1}$, a target radiation tolerance of $\ge$~1~MRad would 
enable its use for approximately one year, which appears to be a valid requirement. 

In this paper we present a prototype CMOS active pixel sensor with radiation tolerant 
pixel cell design for use in TEM. We evaluate its performance applying high energy physics 
(HEP) modelling and experimental techniques to their characterisation as TEM detectors.
We study the particle energy loss, charge spread, point spread function 
and response after irradiation with 200~keV electrons and 29~MeV protons for doses in 
excess to 1~MRad. Data are compared to results of detailed simulation of electron interactions, 
energy loss and signal generation.

\section{Sensor Design and Readout}

\begin{figure}[h!]
\begin{center}
\epsfig{file=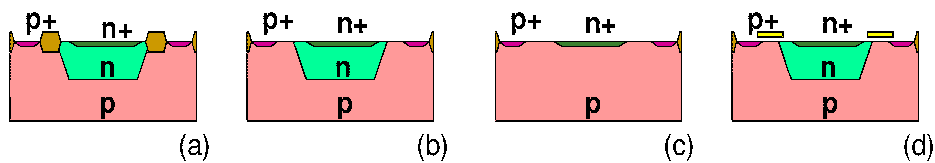,width=12.5cm,clip=}
\end{center}
\caption[]{Cross-section layouts of different diodes used in the test chip. 
The diode in (a), GR, is an n-well diode with an enclosing $p$+ guard ring; 
(b) NW draws thin oxide over the diode; (c) is like (b) but without the well implant; 
and (d) PO is like (b) but with a polysilicon ring covering the potentially inverted region.}
\label{fig:layout}
\end{figure}

We developed a new circuit, named LDRD2-RH, optimised for radiation tolerance, 
as a variation on a previous circuit, developed for HEP applications~\cite{ldrd2}. 
The LDRD2-RH chip retains the original addressing and output circuitry.  The sensor 
is divided into several sectors, each consisting of 24$\times$48 pixels arrayed on a 
20~$\mu$m pitch, in order to test different pixel layouts. The principle radiation 
damage mechanism is radiation-induced leakage 
current in the charge-collecting diode.  This leakage current is primarily caused by the 
positive potential generated by trapped (positive) charge in the oxide above the p-type 
silicon, which results in inversion of the silicon at the
Si/SiO$_2$ interface. In each 20$\times$20~$\mu$m$^2$ pixel, two 
10$\times$20~$\mu$m$^2$ pixels are drawn: each with the same design of the charge-collecting 
diode, and with the classical 3T-readout.  For both sub-pixels, the transistor sizes (W/L) 
are the same, but in one sub-pixel the transistor has a conventional linear layout, 
and in the other, an enclosed gate layout. 
Figure~\ref{fig:layout} shows some of the different diode layouts used.  

The detector has been fabricated in the AMS 0.35~$\mu$m 4-metal, 2-poly CMOS-OPTO 
process, which has an epitaxial layer with a nominal thickness of 14~$\mu$m.

Pixels are read out in rolling shutter mode, which ensures a constant integration 
time across the pixel matrix. 
Pixels are clocked at 6.25~MHz and 25~MHz, corresponding to an integration time 
of 737~$\mu$s and 184~$\mu$s, respectively. The detector is readout through a custom 
FPGA-driven acquisition board. A set of 14~bits, 40~MSample/s ADCs reads the chip outputs, 
while an array of digital buffers drives all the required clocks and synchronisation 
signals. The FPGA has been programmed to generate the clock pattern and collect the 
data sampled by the ADCs. A 32~bit-wide bus connects the FPGA to a National Instruments 
PCI 6533 digital acquisition board installed on the PCI bus of a control PC. Data is 
processed on-line by a LabView-based program, which performs the computation of the 
pixel noise and pedestal levels. Results are given for the PO pixel design, which was 
found to be best performing in terms of radiation tolerance, read out at 6.25~MHz, 
unless otherwise noted.

\section{Response Simulation}

The energy deposition in the sensor active layer and the lateral charge spread 
is simulated using two independent programs.

The simulation of electron scattering in matter for applications in electron 
microscopy has been described in~\cite{DavidJoy}. The first simulation program
is inspired by this formalism, but with the relativistic 
corrections added, as required for accuracy at our energies of interest.
The second simulation is based on the {\tt Geant-4} program~\cite{Agostinelli:2002hh}
and uses the low energy electromagnetic physics models~\cite{Chauvie:2001fh}.

\begin{figure}[h!]
\begin{center}
\begin{tabular}{c c}
\epsfig{file=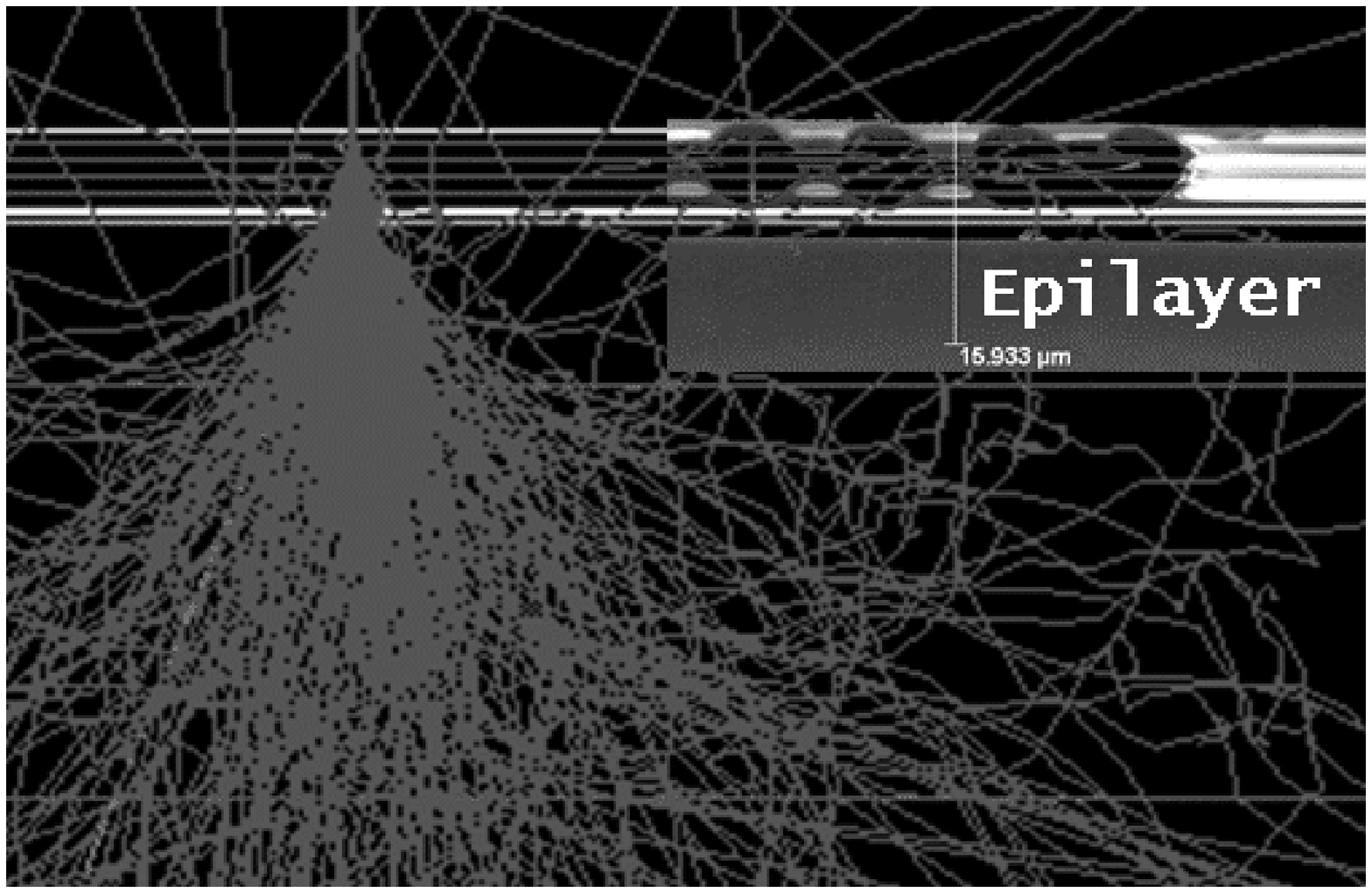,width=6.5cm} &
\epsfig{file=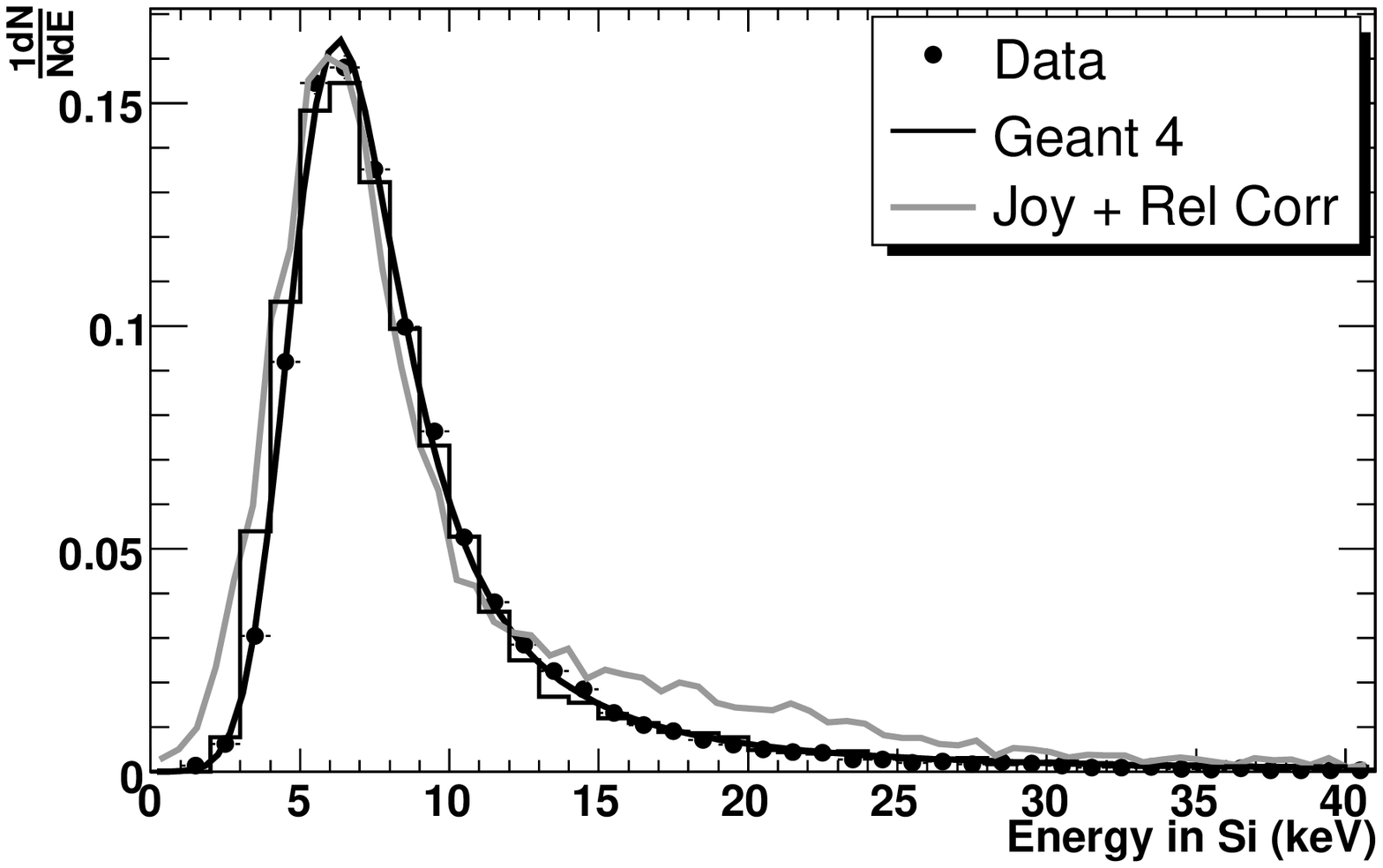,width=7.25cm} \\
\end{tabular}
\end{center}
\caption[]{Response simulation of the LDRD2-RH chip: (Left) Simulated trajectories of 200~keV 
electrons in the pixel sensor. The metal layers on top and the sensitive epitaxial layer below 
are highlighted. (Right) Reconstructed deposited energy for 200~keV electrons. The points with 
error bars show the data, the grey smoothed line the prediction of the simulation according 
to~\cite{DavidJoy} with relativistic corrections added and the histogram the result of the 
{\tt Geant-4} simulation. The continuous line is the fit to the data points with a Landau 
function convoluted with a Gaussian noise term.}
\label{fig:evt}
\end{figure}

In both simulations, the CMOS sensor is modelled according to the detailed 
geometric structure of oxide, metal interconnect and silicon layers, as specified 
by the foundry.
Electrons are incident perpendicular to the detector plane and tracked through the 
sensor. For each interaction within the epitaxial layer, the energy released and the 
position are recorded. Figure~\ref{fig:evt} shows the sensor layout adopted in the 
simulation with the simulated trajectories of 200~keV electrons.

The two simulation programs give a good description of the measured energy deposition as 
shown in Figure~\ref{fig:evt}. {\tt Geant-4} provides with a better description of the high 
energy part of the Landau distribution which tends to be systematically over-estimated by 
the results of the first simulation program. This effect becomes more significant with 
increasing electron energies.
 
The lateral charge spread due to multiple scattering in the sensor is evaluated in 
simulation by computing the lateral position of the electron interaction points in the 
epitaxial layer, weighted by the deposited energy. Results from the {\tt Geant-4} simulation 
are given in Table~\ref{tab:qspread}. 
\begin{table}
\caption[]{Lateral charge spread due to multiple scattering predicted by {\tt Geant-4} 
for various electron energies.}
\begin{center}
\begin{tabular}{|l|c|}
\hline
$E_e$ & Lateral Spread \\ 
      & ($\mu$m) \\ \hline
120~keV & 15.6 \\
160~keV & 11.5 \\
200~keV & 9.3  \\
300~keV & 6.8  \\
1.5~GeV & 1.3  \\ \hline
\end{tabular}
\label{tab:qspread}
\end{center}
\end{table}

Charge collection is simulated with {\tt PixelSim}, a dedicated 
digitisation module~\cite{Battaglia:2007eu}, developed in the {\tt Marlin} 
C++ reconstruction framework~\cite{Gaede:2006pj}. This processor starts from 
the ionisation points generated along the particle trajectory by {\tt Geant-4} 
and models the diffusion of charge carriers from the epitaxial layer to the 
collection diode. {\tt PixelSim} provides us with 
full simulation of the response of all the individual pixels in the detector 
matrix, including electronics noise and efficiency effects, which can be 
processed through the same analysis chain as the data. The simulation has 
a free parameter, the diffusion parameter $\sigma_{{\mathrm{diff}}}$, used to 
determine the width of the charge carrier cloud, which is tuned to reproduce the 
pixel multiplicity in the cluster measured for 1.5~GeV electrons in data, as 
discussed in the next section. We find the best agreement between the simulated and 
measured pixel multiplicity in the cluster for $\sigma_{{\mathrm{diff}}}$ = 16.5~$\mu$m. 
This can be compared with an estimate of the charge diffusion length 
$L_n = \sqrt{D_n \tau_n}$, obtained from the diffusion coefficient, 
$D_n = \frac{kT}{e} \mu_n$, and the charge collection time, $\tau_n$. The diffusion 
coefficient is computed for an estimated doping of $10^{14}$ - $10^{15}$~cm$^{-3}$ 
of the epitaxial layer. The charge collection time has been measured on data. We focus a 
1060~nm laser to a $\simeq$~10~$\mu$m spot onto a single pixel and pulse the 
laser for 2~ns. The pixel analog output is recorded on a digital oscilloscope and we 
observe that the pixel analog level reaches a plateau 150~ns after the arrival of 
the laser pulse. From these data we estimate a diffusion length of 14~$\mu$m - 19~$\mu$m, 
which agrees well with the simulation result.

\section{Sensor Tests}

\subsection{Energy Deposition}

The detector calibration is obtained by recording the position of the 5.9~keV 
full energy peak of a collimated 2.2~mCi $^{55}$Fe source. We find a conversion 
factor of 0.98~keV/ADC count or 26.7~$e^-$/ADC count at 6.25~MHz readout frequency.

Electrons in the energy range from 120~keV to 200~keV from the 200CX electron 
microscope at the National Center for Electron Microscopy (NCEM) are used to 
characterise the detector response to low energy particles. 
The response to high momentum particles is studied with the 1.5~GeV electron 
beam extracted from the LBNL Advanced Light Source (ALS) booster. Data are 
converted into the {\tt lcio} format~\cite{Gaede:2003ip}. Data analysis is 
performed offline by a set of dedicated processors developed in {\tt Marlin}
and proceeds as follows. Events are first scanned for noisy pixels. 
The noise and pedestal values computed on-line are updated, using the algorithm 
in ~\cite{chabaud}, to follow possible variations in the course of data taking.
The measured average pixel noise is (130 $\pm$ 6)~$e^-$ and (71 $\pm$ 4)~$e^-$
of equivalent noise charge (ENC) for the electron microscope and the ALS data 
respectively at 6.25~MHz and (110 $\pm$ 8)~$e^-$ and (67 $\pm$ 5)~$e^-$ at 25~MHz. 
The noise is partly due to the readout electronics, 
and is larger in the electron microscope setup due to the longer cable needed 
to route the analog signals out of the vacuum enclosure at the bottom of the 
microscope column, with a minor contribution from leakage current from operating 
the detector at $\simeq$+25$^{\circ}$C.

Electron hits are then reconstructed from the recorded pixel pulse heights. 
The detector is scanned for pixels with pulse height values over a given 
S/N threshold. These are designated as cluster `seeds'. 
Seeds are sorted according to their pulse heights and the surrounding 
neighbouring pixels are tested for addition to the cluster. The neighbour 
search is performed in a 5$\times$5 matrix around the seed. Pixel thresholds 
at 3.5 and 2.0 units of noise have been used for seed and additional pixels, 
respectively. Clusters are not allowed to 
overlap, i.e. pixels already associated with one cluster are not considered for 
populating another cluster around a different seed. Finally, we require that 
clusters are not discontinuous, i.e.\ pixels associated to a cluster cannot be 
interleaved by any pixel below the neighbour threshold. Reconstructed hits are 
characterised in terms of the energy recorded in a 3$\times$3 pixel matrix centred 
around the seed pixel, the pixel multiplicity of the reconstructed cluster and the 
fraction of the total charge collected by the pixels in the matrix, sorted 
in order of decreasing pulse height. 

\begin{figure}
\begin{center}
\begin{tabular}{c c}
\epsfig{file=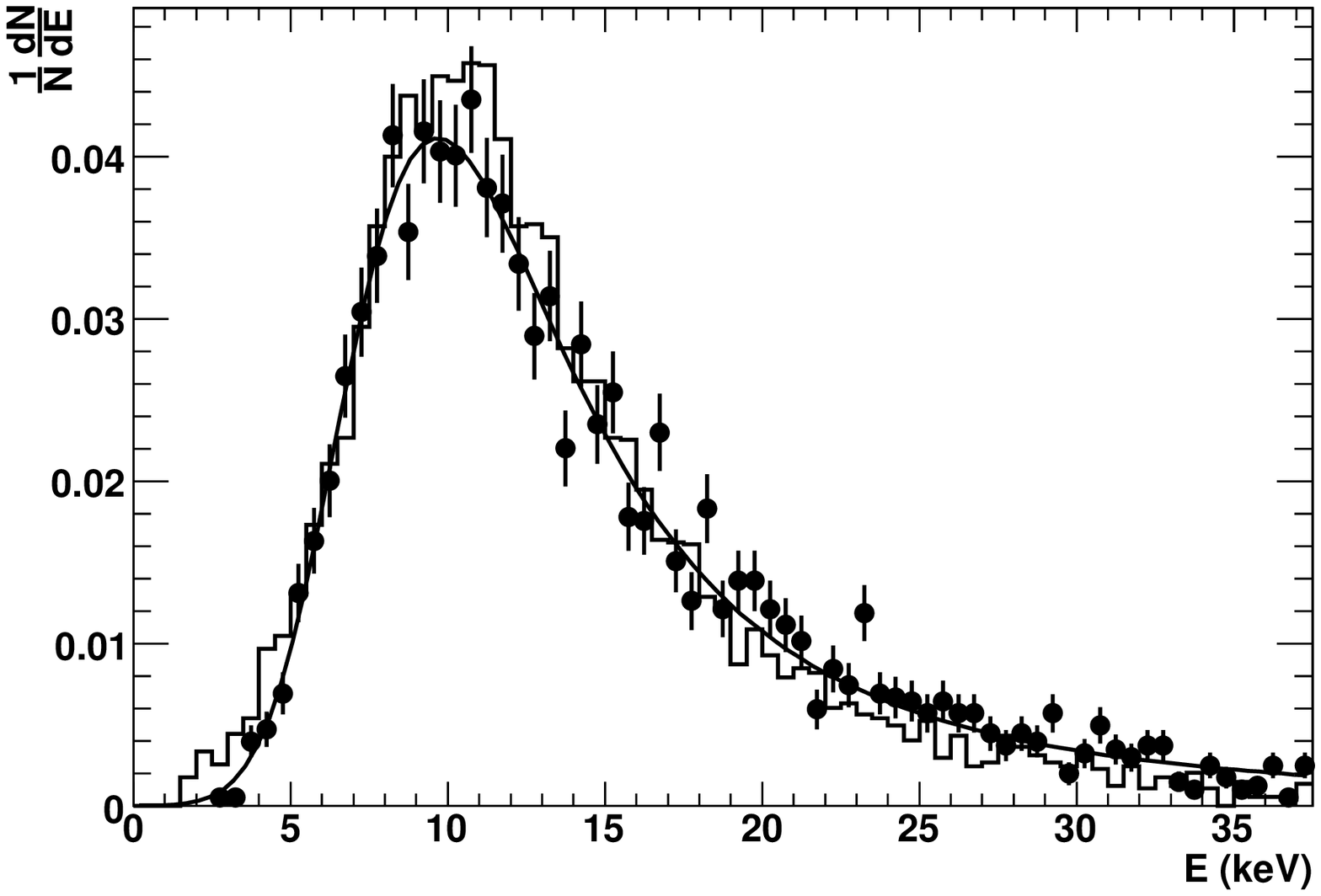,width=7.0cm} &
\epsfig{file=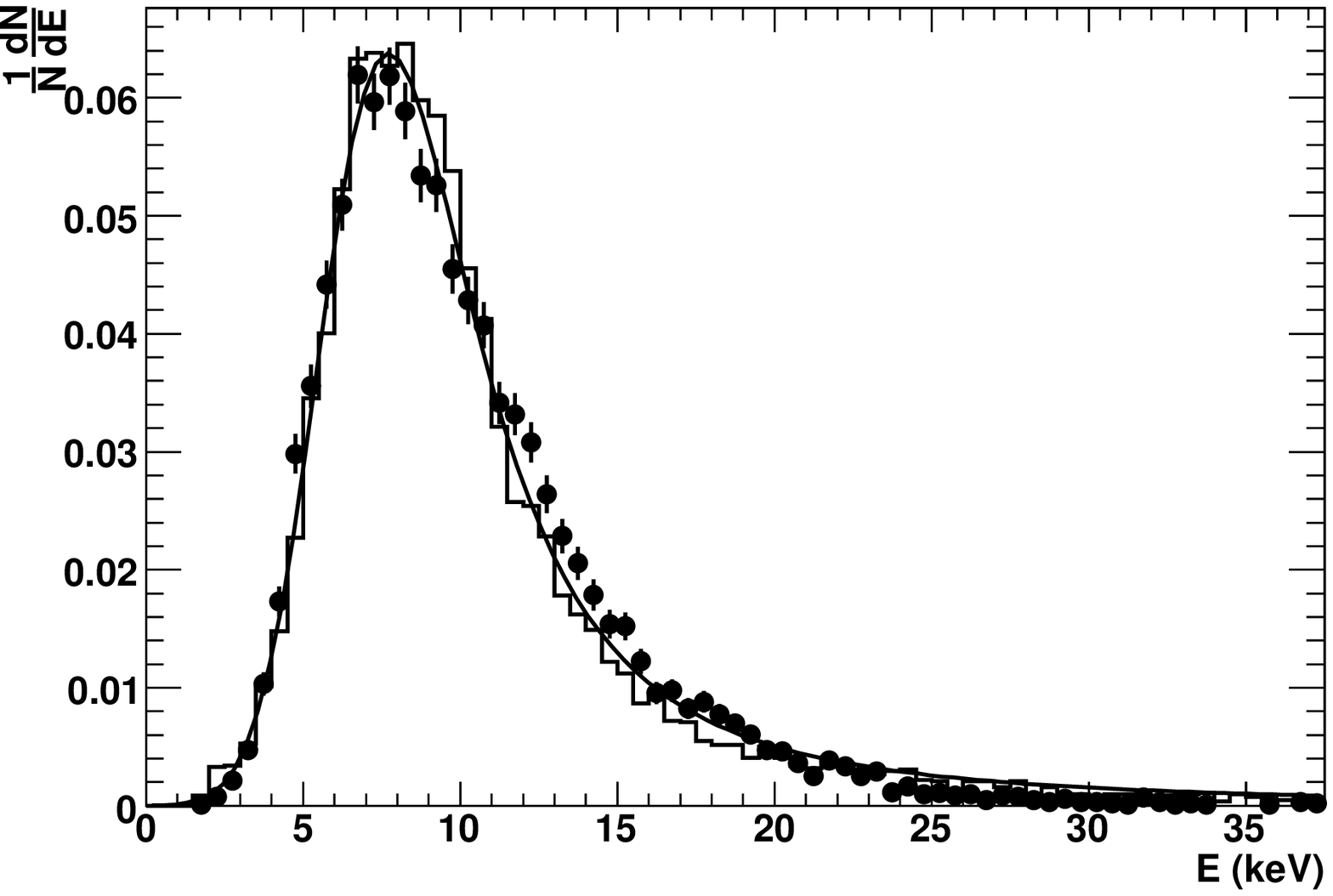,width=7.0cm} \\
\epsfig{file=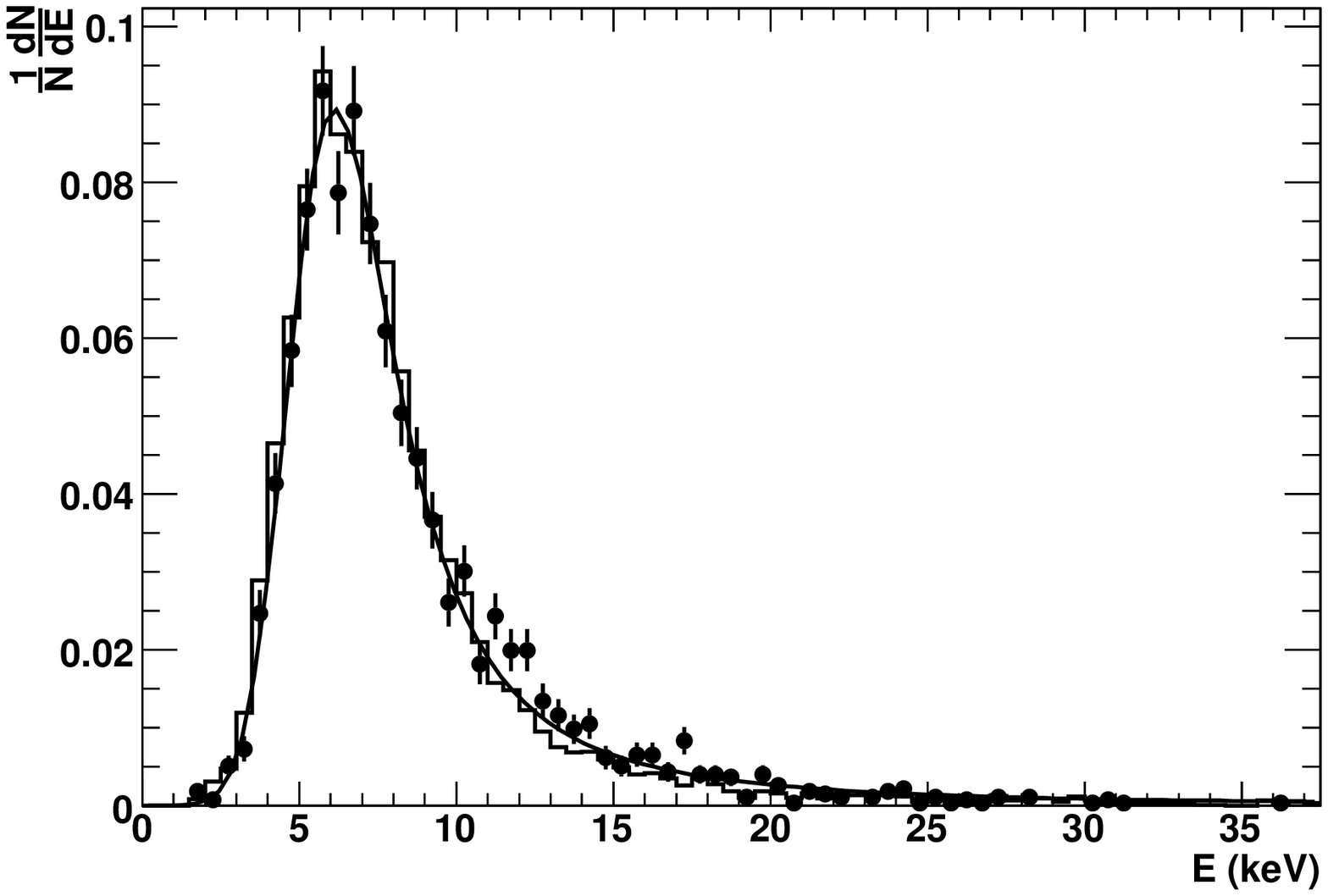,width=7.0cm} &
\epsfig{file=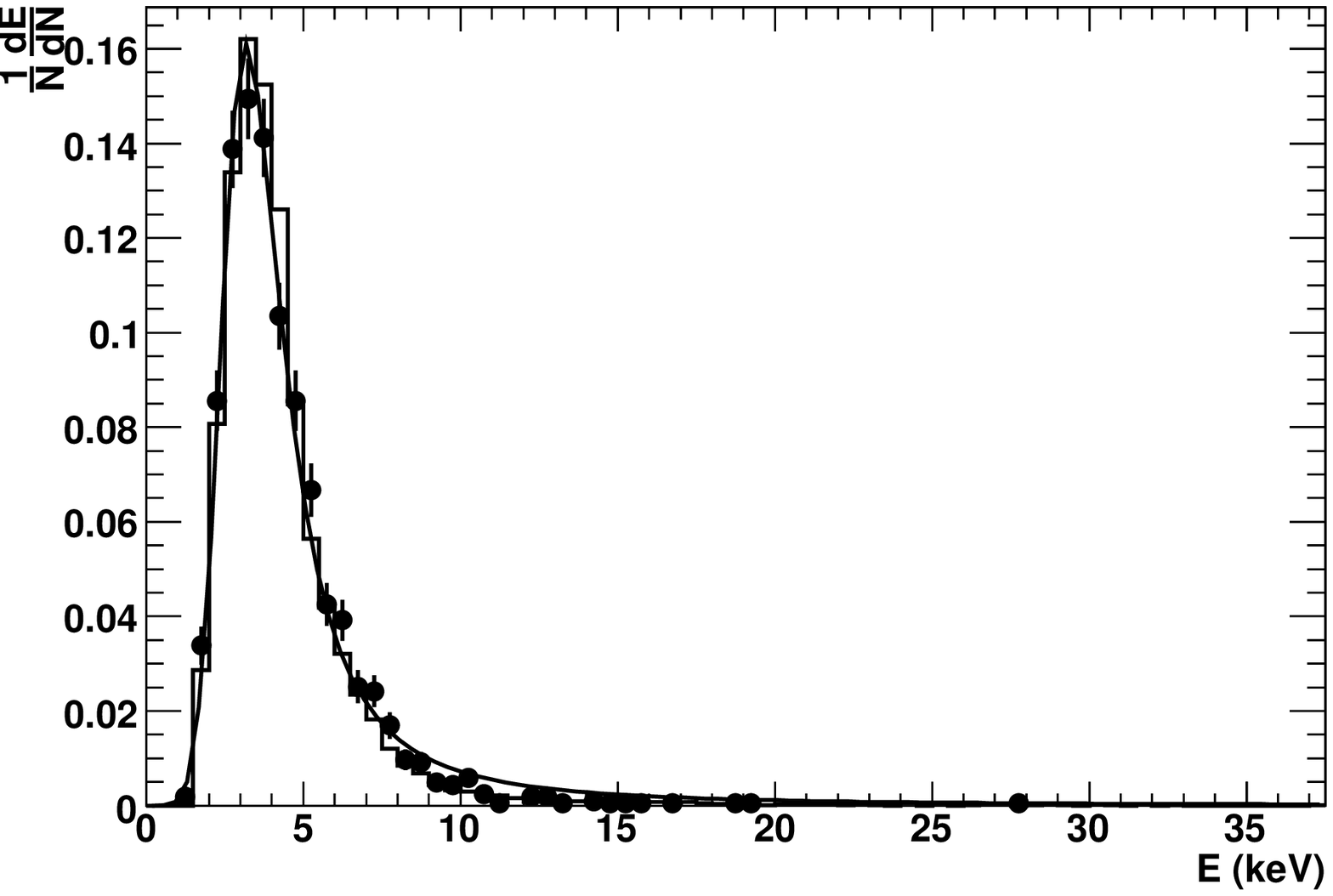,width=7.0cm} \\
\end{tabular}
\end{center}
\caption[]{Reconstructed deposited energy in the 3$\times$3 pixel 
matrix for 120~keV electrons (upper left), 160~keV electrons 
(upper right), 200~keV electrons (lower left) and 1.5~GeV electrons 
(lower right). The points with error bars show the data and the histogram
the result of the {\tt Geant-4} simulation. The continuous line shows
a Landau function convoluted with a Gaussian noise term fit to the data.}
\label{fig:landau}
\end{figure}
Results for the reconstructed deposited energy are shown in Figure~\ref{fig:landau}, 
where data and simulation for 120~keV, 160~keV, 200~keV and 1.5~GeV electrons are 
compared. The simulation reproduces well the measured energy deposition, over this 
range of particle energies. 
The observed average values of energy deposition exceeds those predicted by the 
thin straggling model~\cite{bichsel} for 14~$\mu$m of Si, due to the effect of 
electron interactions in the SiO$_2$ and metal layers on top of the sensitive 
volume.

\subsection{Charge Spread}

The lateral charge spread is studied from the shape of the reconstructed clusters.
The two variables used are the cluster size, i.e.\ the average pixel multiplicity  
$<{\mathrm N_{pixels}}>$, and the distribution of the fraction of the total 
charge collected in the pixels with the highest pulse height in a matrix around 
the seed pixel. Table~\ref{tab:npixels} compares the cluster size for 
electrons of different energies measured in data with the simulation prediction.
\begin{table}
\caption[]{Cluster size for electrons of different energies in data and simulation.}
\begin{center}
\begin{tabular}{|l|c|c|}
\hline
$E_e$ & $<{\mathrm N_{pixels}}>$ & $<{\mathrm N_{pixels}}>$\\ 
(keV) & Data & Simulation \\ \hline
120~keV  & 3.35 $\pm$ 0.06 & 3.01 $\pm$ 0.02 \\
160~keV  & 2.82 $\pm$ 0.04 & 2.73 $\pm$ 0.03 \\
200~keV  & 2.25 $\pm$ 0.04 & 2.33 $\pm$ 0.03 \\ \hline
~1.5~GeV & 2.20 $\pm$ 0.02 & 2.21 $\pm$ 0.02 \\ \hline
\end{tabular}
\end{center}
\label{tab:npixels}
\end{table}
The cluster size for 1.5~GeV electrons is governed by the charge carrier 
diffusion in the epitaxial layer, since the multiple scattering is negligible 
compared to the pixel size, at this energy. Both data and simulation show an 
increase of the multiplicity for lower energies consistent with the combined 
effect of the larger collected charge, which pushes the pulse height on more 
pixels above the S/N threshold for additional pixels, and of the multiple 
scattering in the detector, which displaces the locations of energy deposition 
from the point of entrance of the electron. 
\begin{figure}
\begin{center}
\begin{tabular}{c c}
\epsfig{file=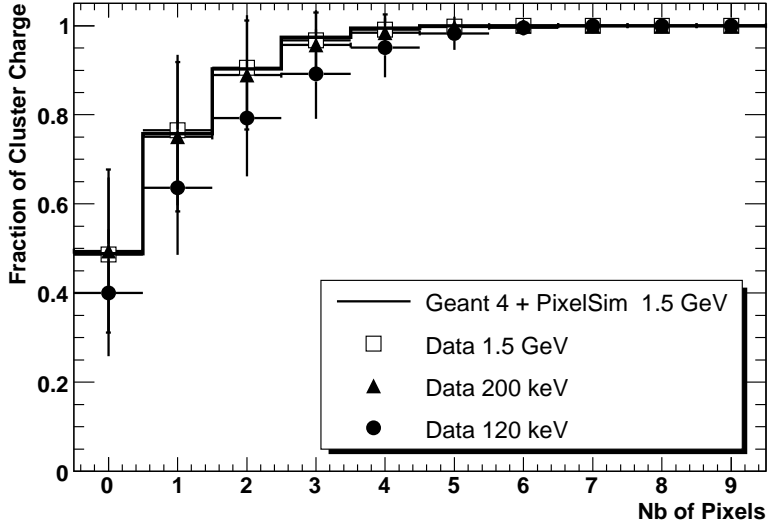,width=11.5cm} &
\end{tabular}
\end{center}
\caption[]{Fraction of the charge collected in a 3$\times$3 pixel matrix as a 
function of the number of pixels added, for 120~keV (dots), 
200~keV (triangles) and 1.5~GeV (squares) data and 1.5~GeV simulation 
(line). The error bars show the r.m.s.\ of the data distribution. Pixels in 
the matrix are sorted in decreasing pulse height order.}
\label{fig:fcharge}
\end{figure}
The two effects can be disentangled by analysing the fraction of the cluster 
charge collected in the pixels with the highest pulse height in a 3$\times$3  
matrix around the cluster seed. Figure~\ref{fig:fcharge} shows the results 
obtained for data at 120~keV and 200~keV compared to 1.5~GeV electrons. 
We indeed observe an increase of the area over which the charge spreads with 
120~keV electrons, consistent with the effect of lateral charge spread due to 
multiple scattering which becomes of the order of the pixel size. On the contrary 
there is no significant increase on the charge distribution for 200~keV electrons, 
where we also observe only a very modest increase in cluster size. These results 
are well reproduced by simulation as shown in Figure~\ref{fig:fcharge}. This first 
validation of simulation in the description of charge spread is important for its 
use in the PSF estimation, which is discussed next.   

\subsection{Point Spread Function and Imaging}

The detector point spread function originates from charge spread due to charge 
carrier diffusion and ionising particle scattering, as well as by the finite 
spatial sampling frequency, which depends on the pixel pitch. We estimate the 
point spread function using {\tt Geant-4} + {\tt PixelSim} simulation and 
test the simulation results with data. 
In simulation, a monochromatic, point-like beam of electrons is sent onto the surface 
of the detector. The point spread function is determined as the r.m.s.\ of the 
predicted distribution of the detected charge on the pixels. Results 
are summarised in Table~\ref{tab:psfsim} for different electron energies and 
10~$\mu$m and 20~$\mu$m pixel pitch. Simulation predicts a PSF better than 10~$\mu$m 
for 300~keV electrons imaged with a 10~$\mu$m pixel pitch.

\begin{table}
\caption[]{Point spread function predicted by Geant 4 + PixelSim for different electron energies 
and 10~$\mu$m and 20~$\mu$m pixel pitch.}
\begin{center}
\begin{tabular}{|l|c|c|}
\hline
Energy & 10~$\mu$m & 20~$\mu$m \\
(keV)  & Pixels    & Pixels    \\ \hline
120    & 11.4      & 13.1      \\
200    & ~9.1      & 11.2      \\
300    & ~8.4      & 10.5      \\ \hline
\end{tabular}
\end{center}
\label{tab:psfsim}
\end{table}

We validate these results using data taken at different energies, from 100~keV up 
to 300~keV, with 10~$\mu$m and 20~$\mu$m pixels at the TITAN test column at NCEM. 
We use the LDRD2-RH as well as an earlier sensor, the LDRD1 chip, featuring simple 
3T pixels with 10~$\mu$m, 20~$\mu$m and 40~$\mu$m pitch and fabricated in the same 
AMS 0.35-OPTO process~\cite{Battaglia:2006ha}. A gold wire was mounted, 
parallel to the pixel columns, at a distance of $\simeq$~3~mm from the detector 
surface. The wire diameter is measured to be (59.6$\pm$0.7)~$\mu$m, using a high 
resolution optical survey system. Since the gold wire has well-defined edges,
the profile of the deposited energy in the pixels, measured across the wire allows 
us to study the charge spread due to scattering and diffusion along the projected 
image of the wire edge and compare to simulation.

\begin{figure}[h!]
\begin{center}
\begin{tabular}{c c}
\epsfig{file=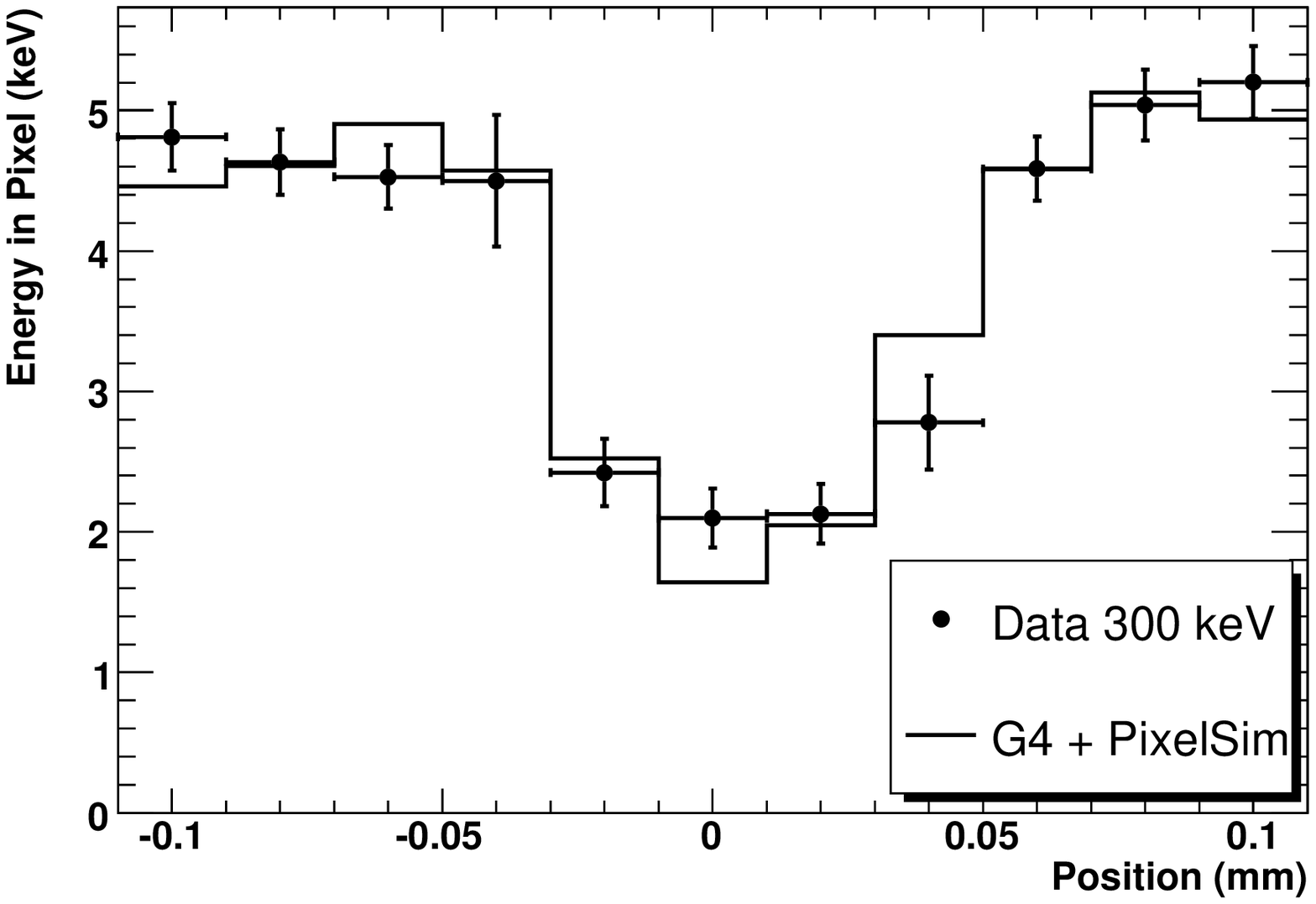,width=7.0cm} & 
\epsfig{file=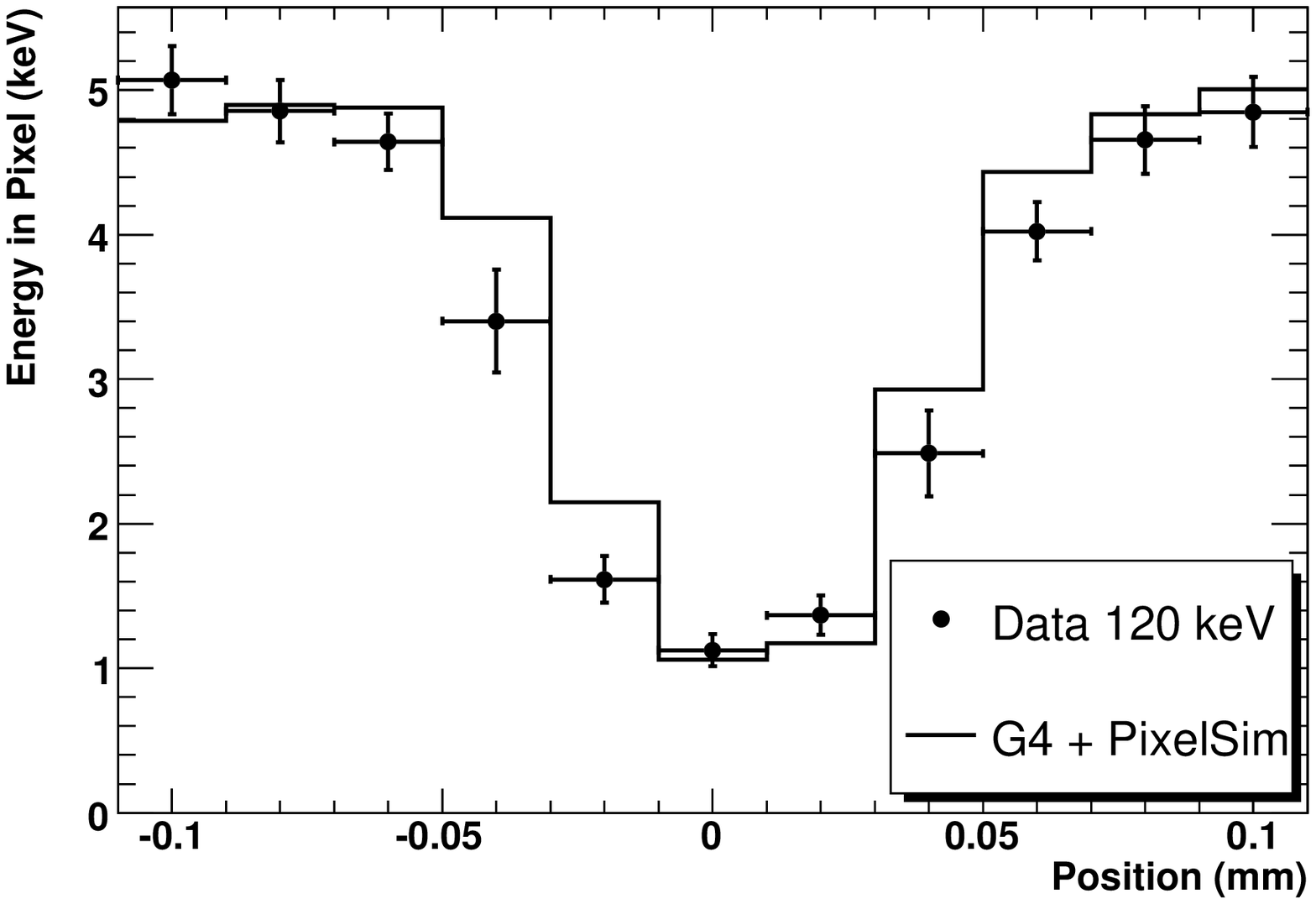,width=7.0cm} \\
\end{tabular}
\end{center}
\caption[]{Pulse heights measured on pixels with 20~$\mu$m pitch, along a row across 
the Au wire stretched above the LDRD2-RH sensor. Data at 300~keV (left) and 120 keV 
(right) are compared with the prediction of the {\tt Geant-4} + {\tt PixelSim} 
simulation.}
\label{fig:wire}
\end{figure}

The setup has been simulated in detail in {\tt Geant-4} and the pixel response 
extracted from {\tt PixelSim}. We study the change in the recorded signal, by
scanning along pixel rows across the gold wire. Figure~\ref{fig:wire} shows 
the pulse heights measured on the pixels along a set of rows, comparing data at 
300~keV and 120~keV with simulation, for the 20~$\mu$m pixel pitch. The good 
agreement observed validates the estimation of the point spread function 
obtained from simulation.

\begin{figure}[h!]
\begin{center}
\begin{tabular}{c}
\epsfig{file=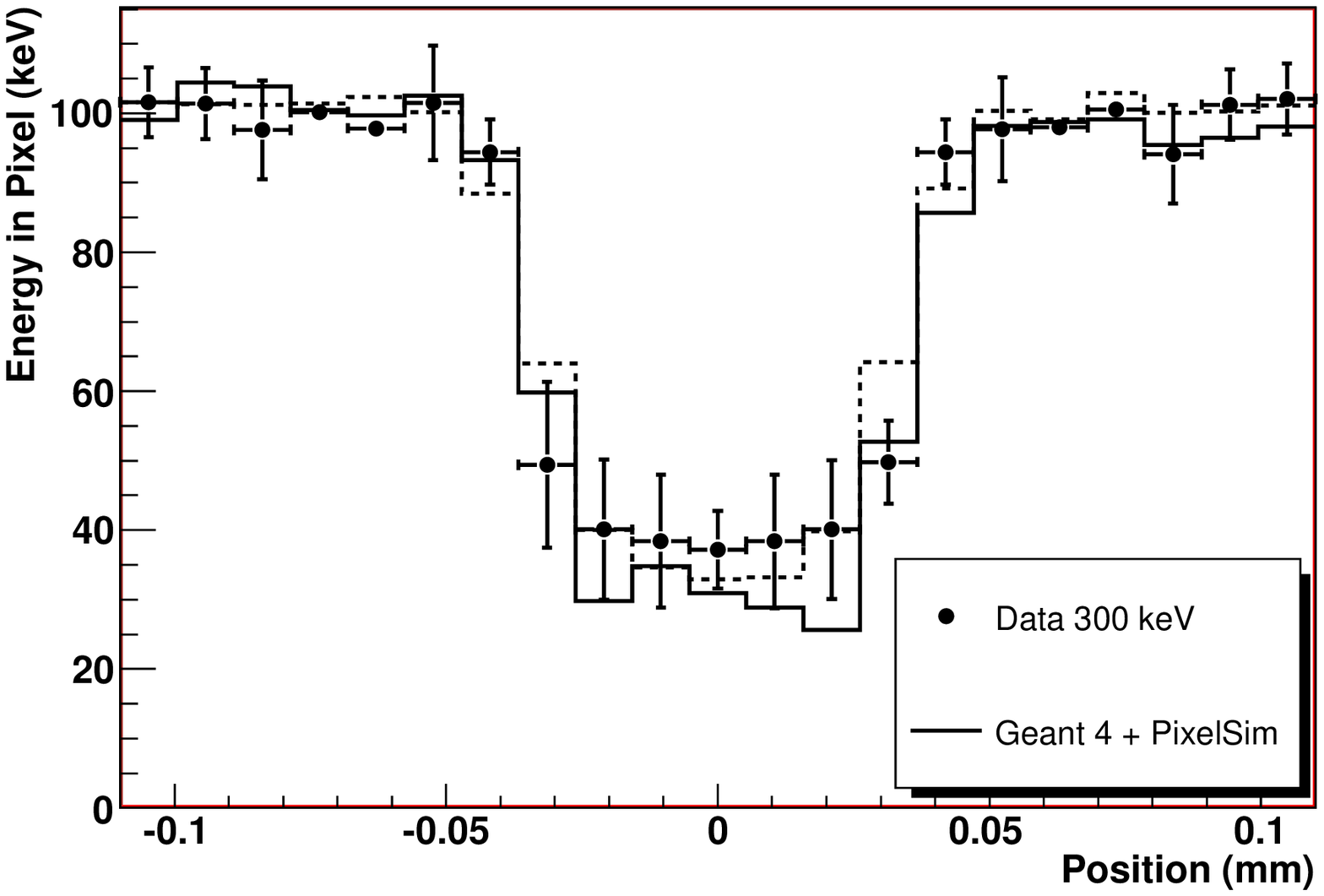,width=11.5cm} \\
\epsfig{file=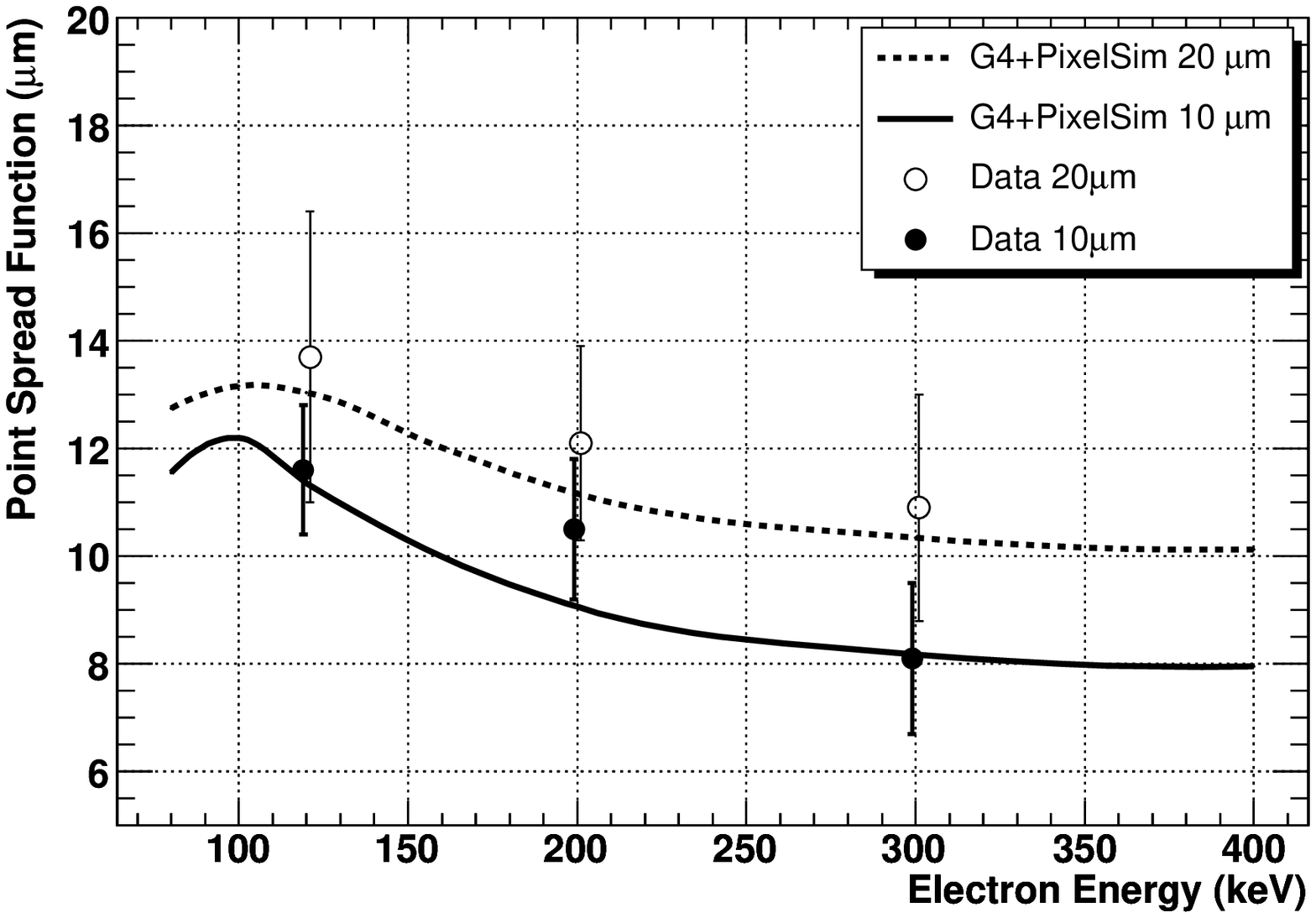,width=11.5cm} \\
\end{tabular}
\end{center}
\caption[]{Point spread function determination. Top: Pulse heights measured on pixels with 
10~$\mu$m pitch, along a row across the Au wire stretched above the 10~$\mu$m pixels of the 
LDRD1 sensor for 300~keV beam. Data (points with error bars) are compared with the prediction 
of the {\tt Geant-4} + {\tt PixelSim} (solid line) simulation and the box+Gaussian function 
fit (dashed line). Bottom: PSF as a function of beam energy. Data (points with 
error bars) are compared to simulation (lines) for 10~$\mu$m and 20~$\mu$m pixels.}
\label{fig:psfit}
\end{figure}
Further, we extract the point spread function directly from the data and compare the 
results to the simulation predictions. We parametrise the measured pulse height on pixel 
rows across the image projected by the wire with a box function smeared by a Gaussian term, 
which describes the point spread function. The maximum and minimum pulse height levels, for 
pixels away from the wire region and for the pixel exactly below the wire centre, respectively, 
are fixed to those observed in data and we perform a simple 1-parameter $\chi^2$ fit to 
extract the Gaussian width term, which gives the PSF. At 300~keV, we measure a PSF value of 
(8.1 $\pm$ 1.6)~$\mu$m for 10~$\mu$m pixels on the LDRD1 chip (see Figure~\ref{fig:psfit}) 
and of (10.9 $\pm$ 2.3)~$\mu$m for 20~$\mu$m pixels on the LDRD2-RH chip, to be compared to
simulation which predicts 8.4~$\mu$m and 10.5~$\mu$m, respectively. The PSF scaling with 
electron energy and pixel size is shown in Figure~\ref{fig:psfit}. Again data and simulation 
are in good agreement. It is interesting to observe that for increasing electron 
energies the point spread function improves, due to the reduction of multiple scattering, 
up to $\simeq$~300~keV, where it approaches an asymptotic value for 10~$\mu$m pixels, 
dominated by the effect of charge spread and spatial sampling frequency. On the opposite 
end of the energy range, the point spread function grows to a maximum around 100~keV 
and simulation predicts it to fall for lower energies, due to the short range in Si of 
these soft electrons.

Finally, we have performed a realistic imaging test with the LDRD2-RH sensor, by 
acquiring images of nanoparticles of a 91.8~\% Pb + 8.2~\% Sn alloy embedded in a 
solid Al matrix~\cite{nanopb} at the 200CX electron microscope. 
The images are obtained with a 200~keV electron flat field illumination at 
6.25~MHz readout frequency and 50~frames~s$^{-1}$ are written to disk. 
We estimate the flux to be $\simeq$~35~$e^-$ pixel$^{-1}$ event$^{-1}$, where one 
event corresponds to a 737~$\mu$s integration time.
\begin{figure}
\begin{center}
\begin{tabular}{c c}
\epsfig{file=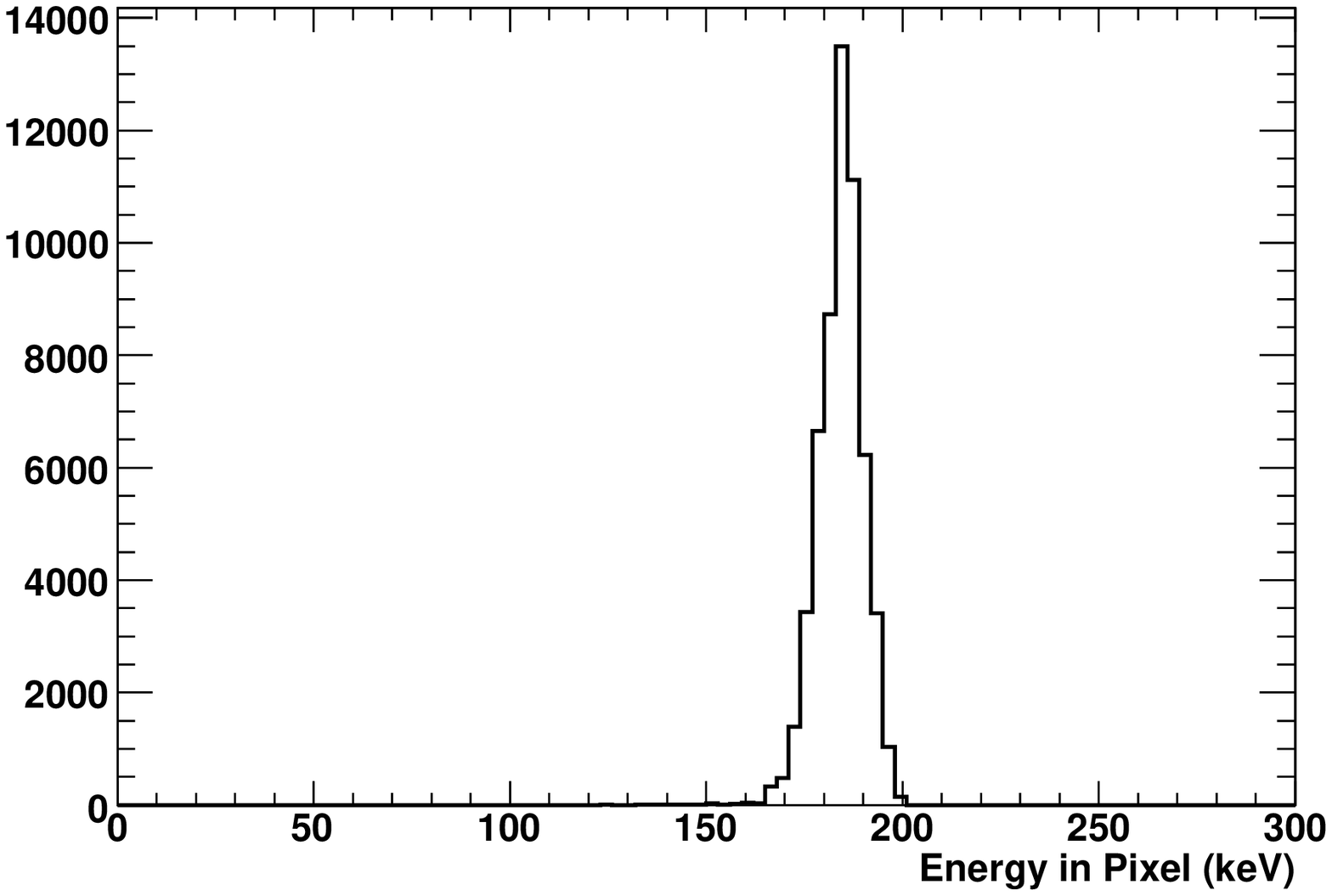,width=7.0cm} &
\epsfig{file=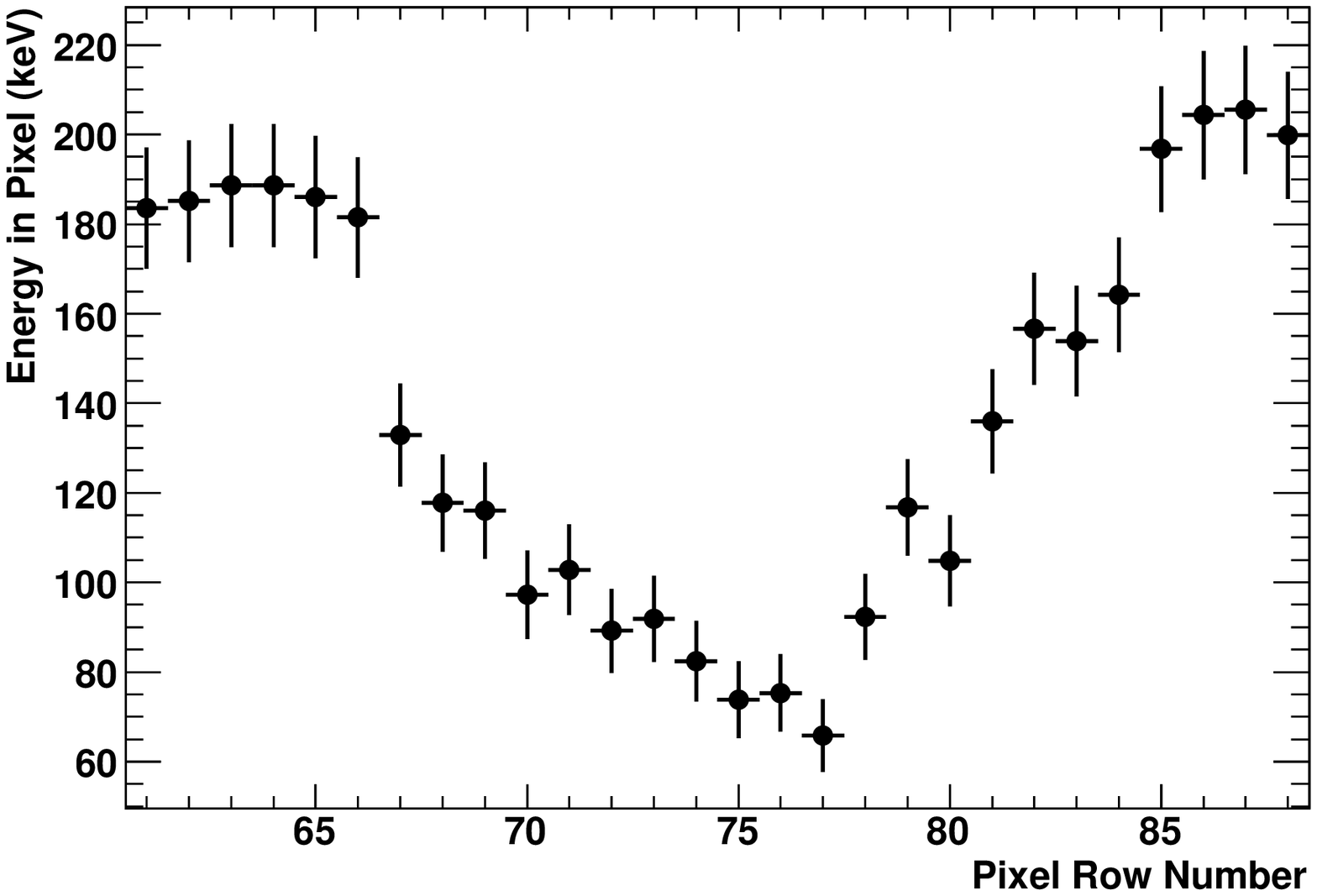,width=7.0cm} \\
\end{tabular}
\end{center}
\caption[]{(Left) Energy recorded on individual pixels with 200 keV electrons flat field 
illumination, (Right) energy recorded on a pixel column across the projected image of a 
Pb-Sn alloy nanoparticle in an Al matrix.}
\label{fig:imaging}
\end{figure}
Figure~\ref{fig:imaging} shows the pulse height recorded on the pixels across the 
projected image of a nanoparticle. 

\section{Sensor Irradiation}

The radiation tolerance of the LDRD2-RH chip for use in TEM has been assessed by comparing 
the sensor response before and after irradiation with 200~keV electrons up to a dose in excess 
of 1~MRad. Results have been integrated by an irradiation with 29~MeV protons at the LBNL 88-inch 
cyclotron. We study the pixel noise, leakage current and charge collection. All tests are performed 
at room temperature.

\begin{figure}
\begin{center}
\begin{tabular}{c c}
\epsfig{file=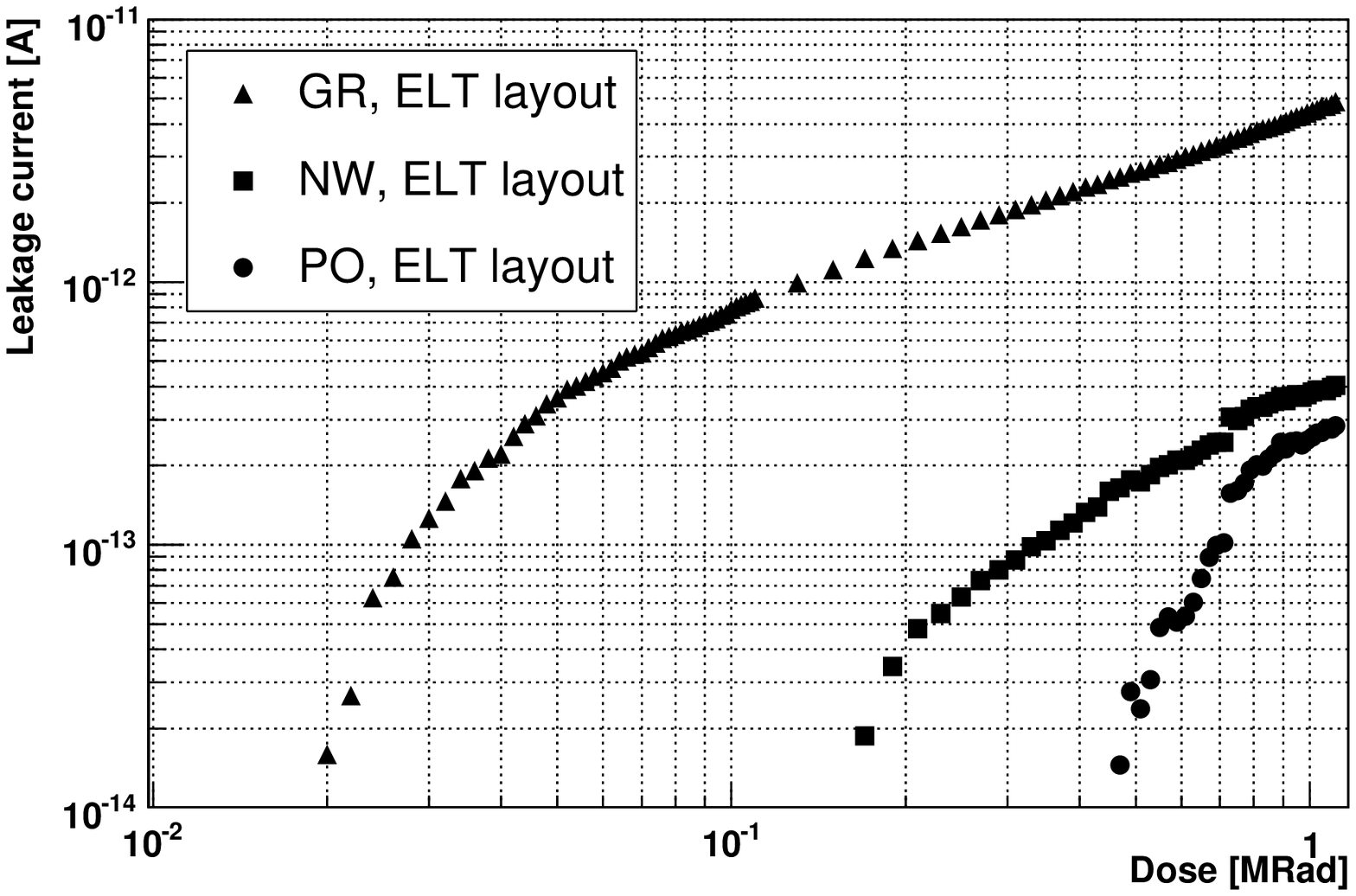 ,width=0.49\columnwidth} &
\epsfig{file=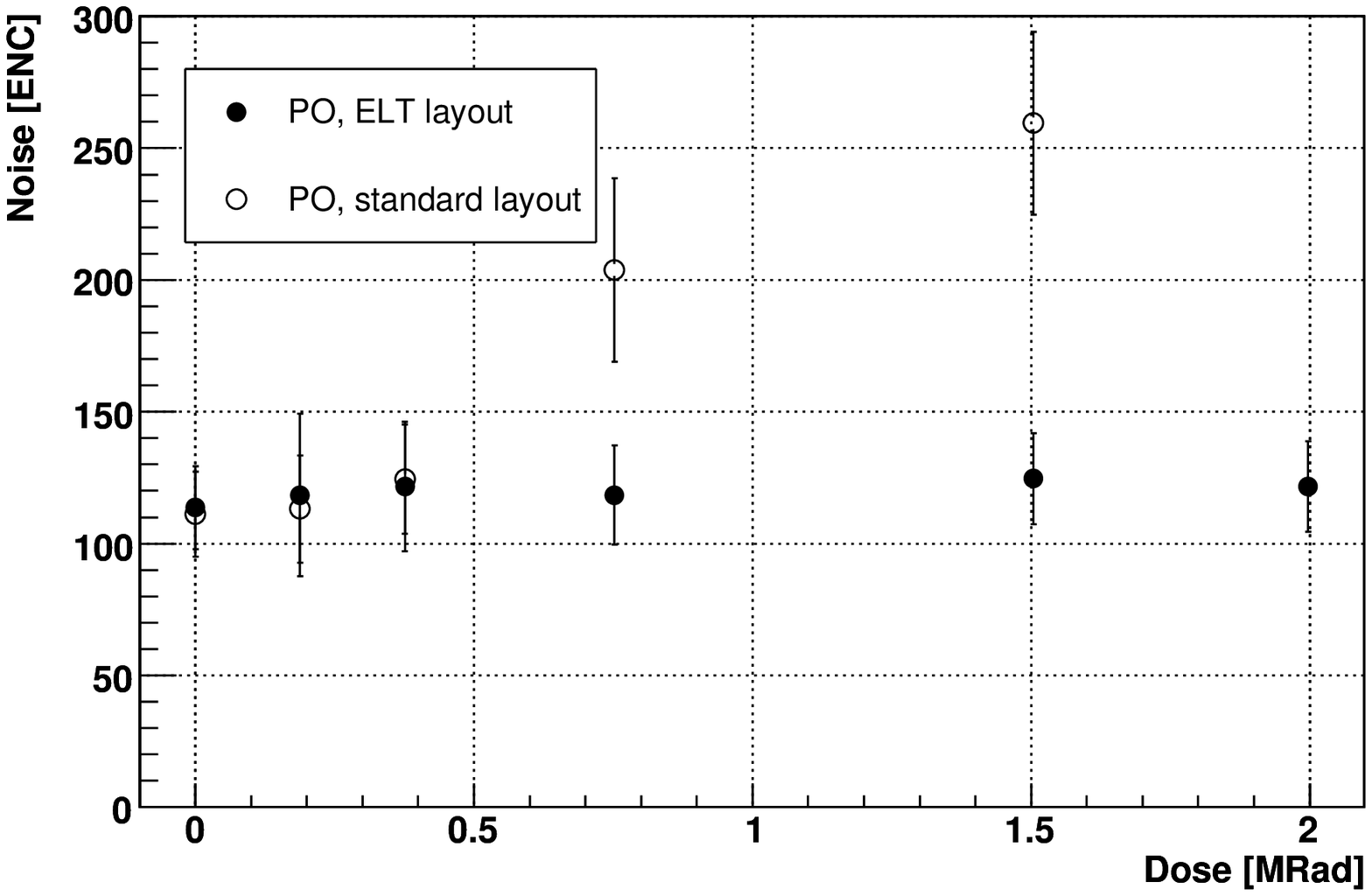,width=0.49\columnwidth} \\
\end{tabular}
\end{center}
\caption[]{Results of the LDRD2-RH sensor irradiation. Left: sensor 
leakage current as a function of dose for 200~keV electrons.
Right: average pixel noise as a function of dose for 29~MeV proton 
in cells designed with ELT and standard layouts. 
The measurement at the highest dose for the standard layout is not
reported since the sensor output signal saturated the ADC dynamic range due to 
the increased pixel leakage current. The error bars represent the r.m.s of the 
pixel noise distributions.}
\label{fig:radpn}
\end{figure}

For the electron irradiation, the sensor active surface has been covered 
with a same gold mesh, having 50~$\mu$m wide bars and 204~$\mu$m 
wide holes. This allows to compare the response of irradiated and non-irradiated 
pixels on the same chip. The sensor has been irradiated with a flux of 
$\simeq$~2300 $e^-$ s$^{-1}$ $\mu$m$^{-2}$, in multiple steps, up to a total 
estimated dose of 1.11~MRad. 
In between consecutive irradiation steps, 100 events are acquired without 
beam and the pixel pedestals and noise computed, in order to monitor 
the evolution of the pixel leakage current with dose. Figure~\ref{fig:radpn} 
shows the increase of the pixel pedestal levels, which measure the leakage current, 
as a function of the integrated dose. 

\begin{figure}
\begin{center}
\begin{tabular}{c c}
\epsfig{file=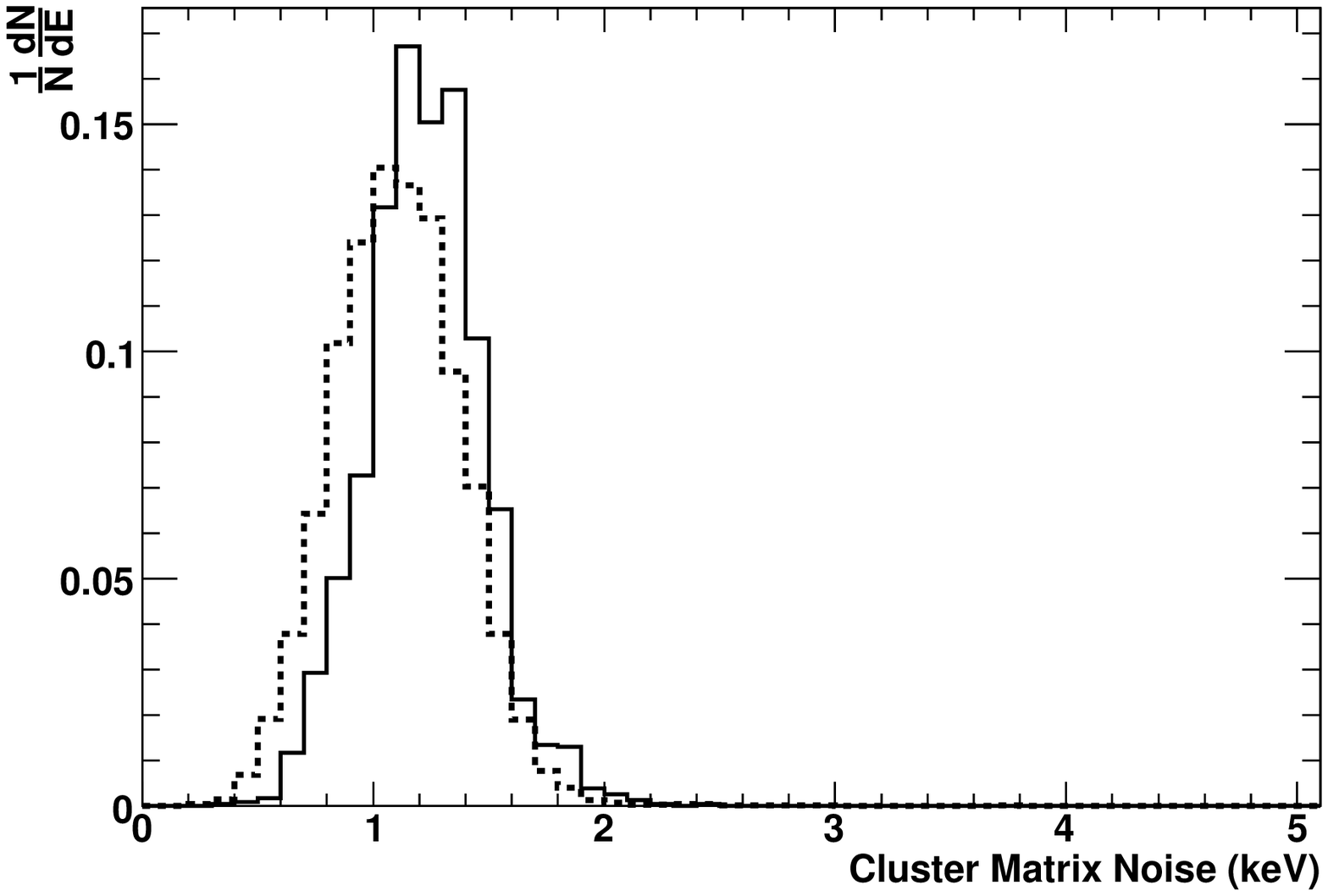,width=7.0cm} &
\epsfig{file=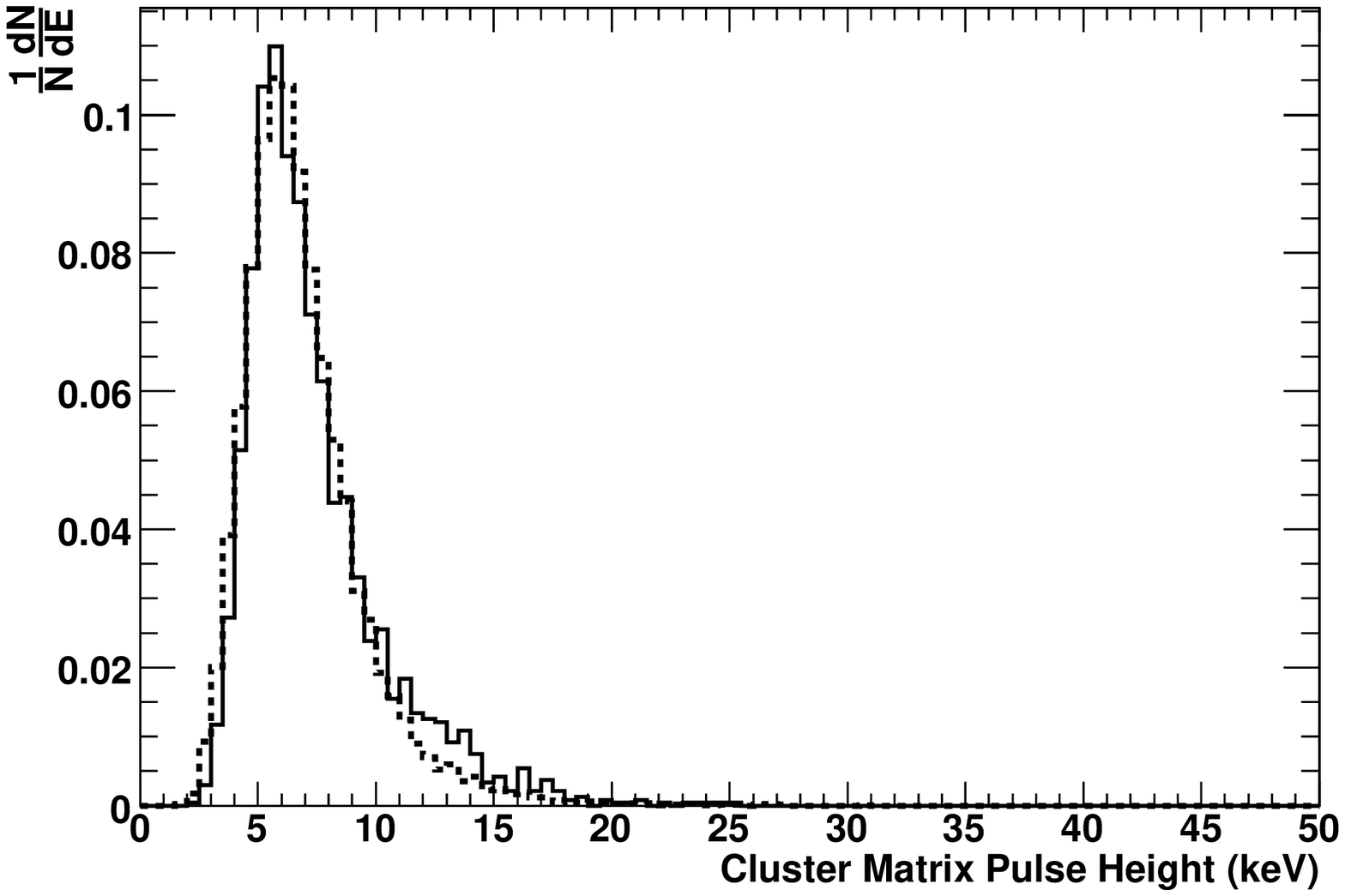,width=7.0cm} \\
\end{tabular}
\end{center}
\caption[]{Measured noise (left) and deposited energy (right) 
in a 3$\times$3 pixel matrix for 200~KeV electrons before 
(continuous line) and after (dashed line) irradiation, obtained 
after applying the gain correction described in the text.}
\label{fig:radcomp}
\end{figure}

After irradiation the sensor response is tested with 200~keV and 1.5~GeV electrons.
The deposited energy in the 3$\times$3 pixel matrix is compared to 
that obtained before irradiation. We fit the energy spectrum with a 
Landau function convoluted with a Gaussian distribution to represent the 
noise contribution. The fit function has three free parameters, the Landau peak 
position (Landau m.p.v.), the Gaussian width (Gaussian Noise) and an overall 
normalisation. Due to the large correlation between the Gaussian width and the 
Landau width, the latter is fixed in the fit to the value obtained on simulation. 
We observe a change of the gain by $\simeq$~1.35, confirmed by calibration with 
$^{55}$Fe. After correcting for this gain shift, the noise and energy deposition 
distributions before and after irradiation are in good agreement. 
In the ALS data, taken with shorter connections between the detector and the 
digitiser board, the electronics noise is lower and an hint of a small increase
of the detector noise after irradiation may be observed. The deposited energy 
spectrum is unchanged after irradiation, confirming that the charge collection 
properties of the pixel cell are not affected. Results are summarised 
in Figure~\ref{fig:radcomp} and Table~\ref{tab:radcomp}. 

\begin{table}
\caption[]{Results of the fits to the distributions of the deposited energy in a 
3$\times$3 pixel matrix for 200~keV and 1.5~GeV electrons before and after the 
1.1~MRad electron irradiation. The noise of the 3$\times$3 pixel matrix is 
also given for comparison. Data after the irradiation is re-calibrated to account 
for the observed gain shift.}
\begin{center}
\begin{tabular}{|l|c|c|}
\hline
               & Before & After \\
               & Irradiation & Irradiation \\ \hline
200~keV $e^-$  &                     &                    \\ \hline
Landau m.p.v.\ & (5.57$\pm$0.04)~keV & (5.54$\pm$0.04)~keV \\
Gaussian Noise & (1.19$\pm$0.04)~keV & (0.98$\pm$0.03)~keV \\ \hline
3$\times$3 Matrix Noise & (1.22$\pm$0.20)~keV & (1.11$\pm$0.20)~keV \\ \hline
1.5~GeV $e^-$  &                     &                    \\ \hline
Landau m.p.v.\ & (3.05$\pm$0.03)~keV & (3.37$\pm$0.07)~keV \\
Gaussian Noise & (0.60$\pm$0.06)~keV & (0.81$\pm$0.07)~keV \\ \hline
3$\times$3 Matrix Noise & (0.74$\pm$0.01)~keV & (0.88$\pm$0.01)~keV \\ \hline
\end{tabular}
\end{center}
\label{tab:radcomp}
\end{table}

Since the threshold energy for electrons
to cause displacement damage in silicon is 260~keV~\cite{lindstroem}, we expect 
200~keV to create only ionising damage. We investigated the effect of 29~MeV protons, 
which damage the sensor via both ionisation and non-ionising 
energy loss. The irradiation has been performed at the 
BASE Facility of the LBNL 88-inch cyclotron~\cite{basef}. Results are reported in details 
in~\cite{radpaper}. One prototype sensor was irradiated with 29~MeV protons up to a 
total integrated fluence of 8.5$\times$10$^{12}$~p/cm$^2$, corresponding to a total dose 
of $\simeq$~2~MRad. At equal doses, a larger increase of the leakage current, compared to 
the electron irradiation, hints at a probable contribution from displacement damage.
Figure~\ref{fig:radpn}~(right) shows the average pixel noise in the PO sectors for ELT 
and linear transistors; it can be seen that while the noise of the ELT layout cells remains 
basically unchanged over the dose range and only slightly increases at the highest doses, 
the noise of the standard layout cells is doubled after $\sim$1~MRad. 
The increased leakage current can be controlled by cooling the chip. The PO design 
with ELT layout appears to be the most radiation tolerant of the designs tested 
in the LDRD2-RH chip.

\section{Conclusions}

Thin monolithic CMOS pixel sensors with small, radiation tolerant pixels 
offer a very attractive solution for fast nano-imaging in transmission 
electron microscopy. The thin sensitive layer ensures a direct detection 
with small multiple scattering contribution to the point spread function. 
Fast readout and single electron sensitivity in small pixel cells result in
high resolution dynamic imaging. We have developed a prototype CMOS pixel 
sensor implementing radiation tolerant cells. The 
sensor response in terms of energy deposition, charge spread and point 
spread function has been extensively tested for electrons in the energy 
range of interest for TEM. A detailed simulation based on {\tt Geant-4} 
and a dedicated charge generation and collection simulation package has 
been validated on the data recorded. 

The point spread function measured with 300~keV electrons is 
(8.1 $\pm$ 1.6)~$\mu$m for 10~$\mu$m pixel and (10.9 $\pm$ 2.3)~$\mu$m 
for 20~$\mu$m pixels, respectively, which agrees well with the values 
of 8.4~$\mu$m and 10.5~$\mu$m predicted by our simulation.

The radiation tolerance of pixels with enclosed transistors and 
specific diode design has been verified with 200~keV electrons and 
29~MeV protons. The pixel cells withstand doses in excess to 1~MRad, 
without significant degradation in noise and charge collection 
efficiency, making them well suited for a sensor for direct detection 
in TEM. 

\section*{Acknowledgements}

\vspace*{-0.1cm}

We wish to thank Thomas Duden, Rolf Erni, Michael Johnson, 
Zhongoon Lee, Peggy McMahan, Marta Rossell Abrodos and the staff 
of the ALS and the 88-inch cyclotron for assistance and for the 
excellent performance of the accelerators.
This work was supported by the Director, Office of Science, of 
the U.S. Department of Energy under Contract No.DE-AC02-05CH11231.

\vspace*{-0.1cm}

 \nolinenumbers

\end{document}